\documentclass[letter,scriptaddress,twocolumn, prl,showpacs,superscriptaddress] {revtex4-1}

\usepackage{graphicx}% Include figure files
\usepackage{dcolumn}% Align table columns on decimal point
\usepackage{bm}% bold math
\begin{document}

\preprint{AIP/123-QED}

\title{Dual gate black phosphorus velocity modulated transistor}

\author{V. Tayari}
\affiliation{Department of Electrical and Computer Engineering, McGill University, Montreal, Qu\'ebec, H3A 0E9, Canada}
\author{N. Hemsworth}
\affiliation{Department of Electrical and Computer Engineering, McGill University, Montreal, Qu\'ebec, H3A 0E9, Canada}
\author{O. Cyr-Choini\`ere}
\affiliation{Department of Physics, McGill University, Montreal, Qu\'ebec, H3A 2T8, Canada}
\author{W. Dickerson}
\affiliation{Department of Electrical and Computer Engineering, McGill University, Montreal, Qu\'ebec, H3A 0E9, Canada}
\author{G. Gervais}
\affiliation{Department of Physics, McGill University, Montreal, Qu\'ebec, H3A 2T8, Canada}
\author{T. Szkopek}
\affiliation{Department of Electrical and Computer Engineering, McGill University, Montreal, Qu\'ebec, H3A 0E9, Canada}
\date{\today}

\maketitle

The layered semiconductor black phosphorus has attracted attention as a 2D atomic crystal that can be prepared in ultra-thin layers for operation as field effect transistors \cite{Li_NN2014, Xia_NC2014, Liu_ACN2014}. Despite the susceptibility of black phosphorus to photo-oxidation \cite{Favron_NM2015}, improvements to the electronic quality of black phosphorus devices has culminated in the observation of the quantum Hall effect \cite{Li_QHEbP}. In this work, we demonstrate the room temperature operation of a dual gated black phosphorus transistor operating as a velocity modulated transistor \cite{Sakaki}, whereby modification of hole density distribution within a black phosphorus quantum well leads to a two-fold modulation of hole mobility. Simultaneous modulation of Schottky barrier resistance leads to a four-fold modulation of transconductance at a fixed hole density. Our work explicitly demonstrates the critical role of charge density distribution upon charge carrier transport within 2D atomic crystals. 

Black phosphorus (bP) is an elemental allotrope and a direct bandgap semiconductor with a puckered, honeycomb layer structure \cite{keyes,Morita} that can be exfoliated down to atomic few-layer thickness \cite{Li_NN2014, Xia_NC2014, Liu_ACN2014, Favron_NM2015, Koenig_APL2014, Gomez_2DM2014}. Although bP is the most thermodynamically stable allotrope of phosphorus, photo-oxidation in the presence of water, oxygen and visible light is known to degrade bP with a reaction rate that increases as bP layer thickness decreases \cite{Favron_NM2015}. Several materials have been used to encapsulate bP in order to protect it against photo-oxidation, including hexagonal boron-nitride \cite{Li_NN2015, Gillgren_2DM2015, Chen_NC2015}, aluminum oxide \cite{Wood}, parylene \cite{Favron_NM2015}, and poly-methylmethacrylate \cite{Tayari_NC2015}. Recent works have also shown that 2D hole transport can be achieved in a single 2D sub-band within an accumulation layer of many-layer bP \cite{Tayari_NC2015, Li_NN2015, Chen_NC2015}, effectively combining 2D transport characteristics with the increased chemical stability of many-layer bP. These advances have culminated in the observation of the quantum Hall effect in bP \cite{Li_QHEbP}. Nonetheless, further understanding and control of transconductance, carrier mobility and contact resistance in bP field effect transistors (FETs) is desired.

We report here an experimental investigation of the transport characteristics of bP FETs with an asymmetric dual gate geometry consisting of top and bottom gate electrodes. The top gate is found to be effective in modulating the back gate FET transfer characteristics, including both field effect mobility and Schottky barrier contact resistance. The mobility modulation effect enables operation of the dual gate bP FET as a velocity modulated transistor (VMT), first proposed by Sakaki \cite{Sakaki} to overcome the limitation on transistor switching frequency imposed by the channel transit time of charge carriers. Mobility modulation has since been demonstrated in GaAs/AlGaAs heterojunctions \cite{Hirakawa}, wide GaAs/AlGaAs quantum wells \cite{Kurobe}, silicon-on-insulator FETs \cite{Prunnila} and the LaAlO$_3$/SrTiO$_3$ interface \cite{Bell}. Room temperature VMT operation in silicon-on-insulator FETs has been demonstrated with up to 1.4-fold mobility modulation \cite{Prunnila}. Asymmetric dual-gate bP FETs exhibit a two-fold mobility modulation at room temperature, and the underlying mechanism is modulation of hole density distribution with the naked bP quantum well channel of the bP FET, and a resultant modulation of scattering by charged impurities within the gate oxide, surface roughness, and other spatially dependent scattering mechanisms. Simultaneously, bP FETs exhibit strong Schottky barrier modulation. First conclusively observed in carbon nanotube FETs \cite{Heinze}, Schottky barrier modulation has recently been shown to dominate off-state conductance of bP FETs \cite{Penumatcha}. The combined effects of mobility modulation and Schottky barrier modulation of dual-gate bP FETs enables four-fold transconductance modulation at a fixed carrier density of $4\times10^{11}$cm$^{-2}$.

\begin{figure}
    \includegraphics [width=2.75in]{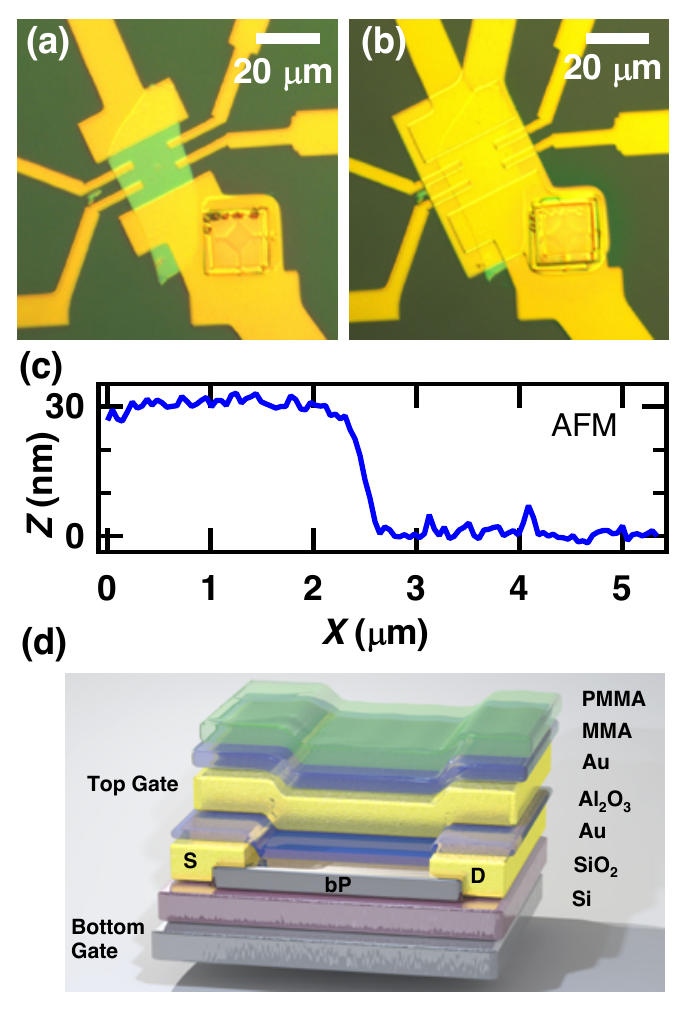}
        \caption{ a) Optical image of a bP FET in a multiple terminal geometry prior to top gate deposition. Scale bar is 20 $\mu$m. b) Optical image of the device shown in b) including top gate and MMA/PMMA encapsulation. Scale bar is 20 $\mu$m. c) Atomic force microscope (AFM) line scan of a region of bare bP/Al$_2$O$_3$ of the device in b), with the MMA/PMMA layers removed. d) Schematic view of a dual gate bP FET with SiO$_{2}$ back-gate, Al$_{2}$O$_{3}$ top gate dielectric, source (S) and drain (D) contacts, and encapsulating layers of MMA and PMMA.  }
        \label{Fig1}
\end{figure}

Nanometer-scale bP crystals were exfoliated using a polydimethylsiloxane (PDMS) stamp technique in a glove box environment \cite{Tayari_NC2015}, and transferred to degenerately doped Si/SiO$_{2}$ wafers functionalized with hexamethyldisilazane (HMDS) layer. The hydrophobic HMDS layer aids in protecting the freshly cleaved surface of the bP from water adsorbates on the SiO$_{2}$ surface, and suppresses charge transfer doping that would otherwise lead to hysteresis and instability in FET characteristics \cite{Lafkioti_NL2010}. Further micro-fabrication was performed to define contact electrodes, a top gate structure, and final encapsulation. An optical image of a typical bP FET in a multiple terminal geometry is shown in Fig.~\ref{Fig1}(a) prior to top gate fabrication and in Fig.~\ref{Fig1}(b) after top gate fabrication. The thickness of the bP layer under the top-gate of the bP FET was determined to be 32~nm by atomic force microscopy, as shown in Fig.~\ref{Fig1}(c). A schematic of the complete bP FET structure is displayed in Fig.~\ref{Fig1}(d). Encapsulating the bP layer between an HMDS functionalized SiO$_2$/Si substrate and an optically opaque gate stack was found to effectively mitigate degradation due to photo-oxidation. The $I-V$ characteristics of our bP FETs were stable over a period of six months.

\begin{figure}
    \includegraphics [width=2.75in]{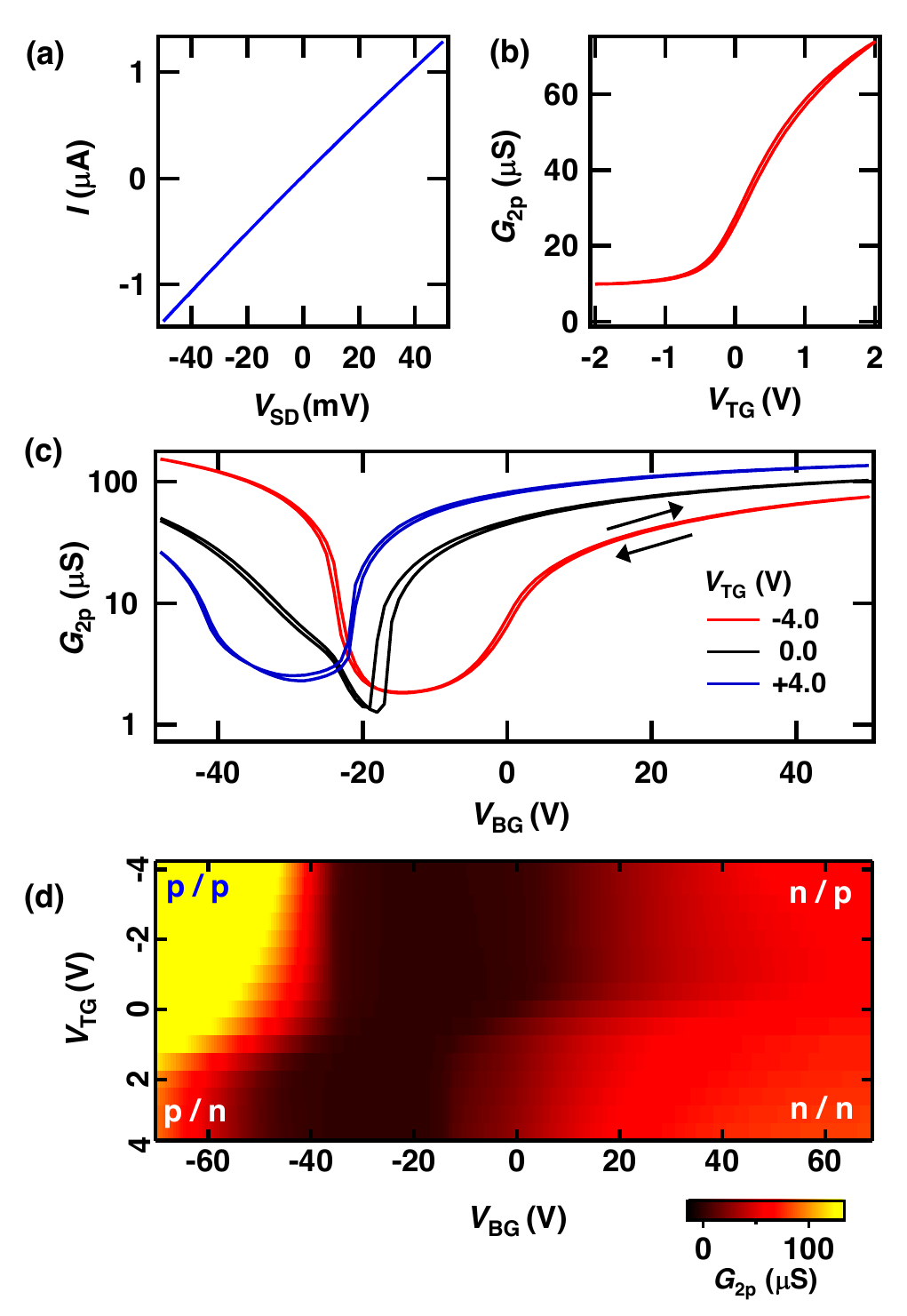}
        \caption{ a) Source-drain current $I$ as a function of the source-drain voltage $V_{SD}$ with back gate and top gate biases of $V_{TG}=0$~V $V_{BG}=0$~V, indicative of Ohmic contacts. b) Two-terminal conductance $G_{2p}$ as a function of top gate voltage $V_{TG}$ at a fixed bottom gate voltage $V_{BG}=0$~V. Electron conduction is observed. c) Two-terminal conductance $G_{2p}$ versus back gate voltage $V_{BG}$ at three top gate voltages $V_{TG} = -4, 0, +4$~V at room temperature. Minimal hysteresis is observed at the 1~V/s sweep rate used for the back gate potential. d) Two dimensional contour plot of two-terminal conductance $G_{2p}$ versus gate voltages $V_{TG}$ and $V_{BG}$ at $T=77$~K, for the same device. }
        \label{Fig2}
\end{figure}

Quasi-dc charge transport measurements were performed at room temperature. Fig.~\ref{Fig2} shows the $I-V$ characteristics of the device shown in Fig.~\ref{Fig1}. The measured two-terminal source-drain current $I$ versus source-drain bias voltage $V_{SD}$ is plotted in the Fig.~\ref{Fig2}(a) with 0~V applied to the top and bottom gates. The linear $I-V_{SD}$ characteristic indicates ohmic, or quasi-ohmic, behaviour of the contact electrodes. The two-terminal conductance $G_{2p}$ as a function of top gate voltage $V_{TG}$ is plotted in Fig.~\ref{Fig2} (b), demonstrating strong conductance modulation consistent with electron conduction and a negligible hysteresis. Gate leakage currents were recorded simultaneously in all of our charge transport experiments, never exceeding $5\%$ of the source-drain current and generally being much lower than the source-drain current. The two-terminal conductance $G_{2p}$ at a constant bias current of 4 $\mu$A as a function of back gate voltage $V_{BG}$ is plotted in Fig.~\ref{Fig2} (c) with top gate voltage held at $V_{TG}$=-4V, 0V and +4V. The room temperature conductance modulation reaches two orders of magnitude, and there is minimal hysteresis in conductance as back gate voltage is swept at a rate of $\pm$1~V/s, which we attribute to the HMDS functionalization of the oxide layer below the bP. The threshold voltages for the onset of electron and hole conduction is modulated by the applied top gate potential. An increasingly negative top gate voltage results in increased back gate threshold voltages for both electron and hole conduction, as expected.

We investigated the dependence of the bP FET conductance as a function of both top and bottom gate voltages. The measured two-point conductance $G_{2p}$ is plotted in Fig.~\ref{Fig2}(d) as a colour contour versus both $V_{TG}$ and $V_{BG}$. An insulating region (dark) is visible in the contour plot, corresponding to minimal mobile carrier density within the bP channel. With $V_{BG}<0$~V and $V_{TG}<0$~V, both gate potentials induce holes within the bP to result in strong hole conduction, identified as p~/~p in Fig.~\ref{Fig2}(d). In contrast, with $V_{BG}>0$~V and $V_{TG}>0$~V, both gate potentials induce electrons  and electron conduction is unambiguously observed, identified as n~/~n. The top gate voltage can also be used to induce opposite carrier type to that induced by the bottom gate,  identified as p~/~n and n~/~p in Fig.~\ref{Fig2}(d).

The top gate potential influences the back gate threshold voltage for both hole and electron conduction over a narrow range -2~V$<V_{TG}<$2~V, beyond which the top gate voltage has comparatively little influence upon channel conductance. The inability of the top gate to induce electron or hole conduction over the back gate voltage range -40~V$<V_{BG}<$-10~V indicates that the charge carriers induced by the top gate are of very low mobility and may to a large extent be localized at charge traps. The Al$_2$O$_3$ atomic layer deposition process takes place under strongly oxidative conditions that may be responsible for the introduction of charge traps and scattering centres at the top bP surface. The asymmetric behaviour of the asymmetric dual gate bP FET is distinct from the symmetric behaviour of symmetric dual gate bP FETs \cite{Kim_NL2015}.

\begin{figure}
   \includegraphics [width=2.75in]{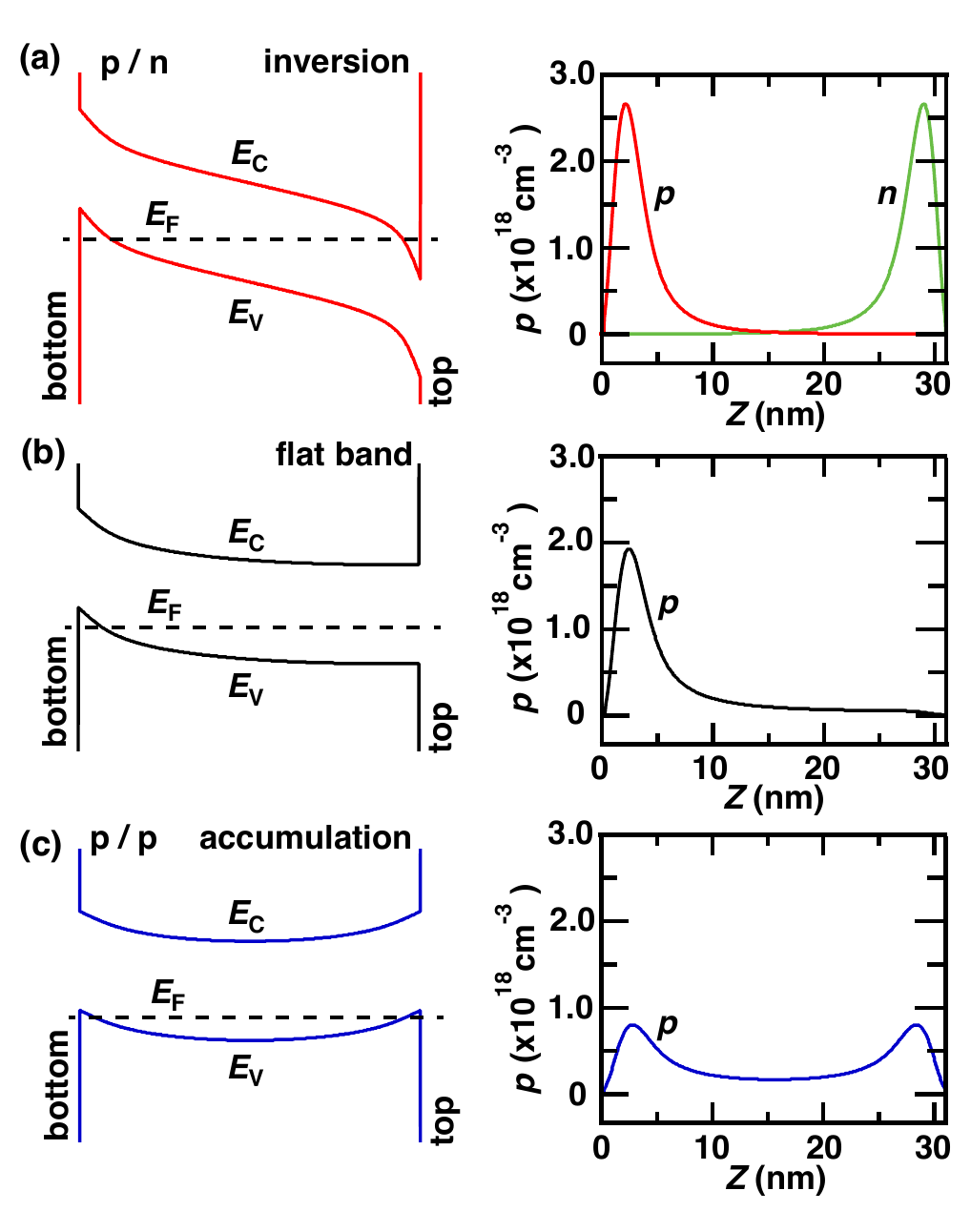}
       \caption{ The band diagrams and volumetric charge density distribution in a 32 nm wide bP quantum well determined by Schr\"{o}dinger-Poisson calculations at $T=300$~K for different gate bias potentials. a) Asymmetric gate bias inducing a p / n carrier distribution with $p = n =10^{12}\mathrm{cm}^{-2}$ and an associated inversion in carrier type. b) Gate bias inducing $p = 10^{12}\mathrm{cm}^{-2}$ at one bP surface and flat-band conditions at the other. c) Symmetric gate bias inducing a p / p carrier distribution with a total hole concentration of $p = 10^{12}\mathrm{cm}^{-2}$, corresponding to hole accumulation at both bP surfaces. }
       \label{Fig3}
\end{figure}

Self-consistent Schr\"{o}dinger-Poisson calculations combining an effective mass theory for bP and a mean-field approximation to Coulomb interactions were employed to gain further insight into the behaviour of the dual gated bP FET. Effective masses for bulk bP determined by cyclotron resonance experiments \cite{narita} were used in our calculations. The band diagram and volumetric hole density with the bP layer are shown in Fig.~\ref{Fig3}(a) at $T$=300~K for a negative back gate voltage and positive top gate voltage adjusted to induce a total hole density of $p=10^{12}\mathrm{cm}^{-2}$ and a total electron density $n=10^{12}\mathrm{cm}^{-2}$. Under these bias conditions, a p / n junction is formed vertically within the bP layer, with holes (electrons) confined at the bottom (top) of the bP. If conduction is strongly suppressed at the top surface due to ALD processing, hole conduction will dominate. Moreover, the top gate potential is screened by the electrons within the inversion layer at the top surface, as observed in our experimental data with $V_{TG}>$2~V. The band diagram and volumetric hole density are shown in Fig.~\ref{Fig3}(b) at $T$=300~K with gate voltages adjusted to induce a total hole density of $p=10^{12}\mathrm{cm}^{-2}$ and flat bands at the top surface. The holes are less tightly confined to the bottom of the bP layer under these conditions. The top gate voltage is no longer screened by induced electrons, and will therefore modulate the threshold back gate voltage for the onset of hole conduction, as observed in our experiments for -2~V$<V_{TG}<$2~V. The band diagram and volumetric hole density with the bP layer are shown in Fig.~\ref{Fig3}(c) at $T$=300~K for negative back gate and top gate voltages adjusted to induce a total hole density of $p=10^{12}\mathrm{cm}^{-2}$ distributed symmetrically within the structure. The top gate potential is screened from influencing the hole density at the bottom of the bP layer, and the volumetric hole density extends within the bulk of the bP layer. At the hole density $p=10^{12}\mathrm{cm}^{-2}$ used for our calculations, the Fermi temperature $T_F = p / ( k_B m^*/  \pi \hbar^2 ) = 126$~K for holes accumulating within a single 2D sub-band. Analysis of 2D sub-band population reveals that two 2D sub-bands are substantially populated for the carrier densities accessed in our calculations, leading to non-degenerate carrier statistics at room temperature.

\begin{figure}
   \includegraphics [width=2.75in]{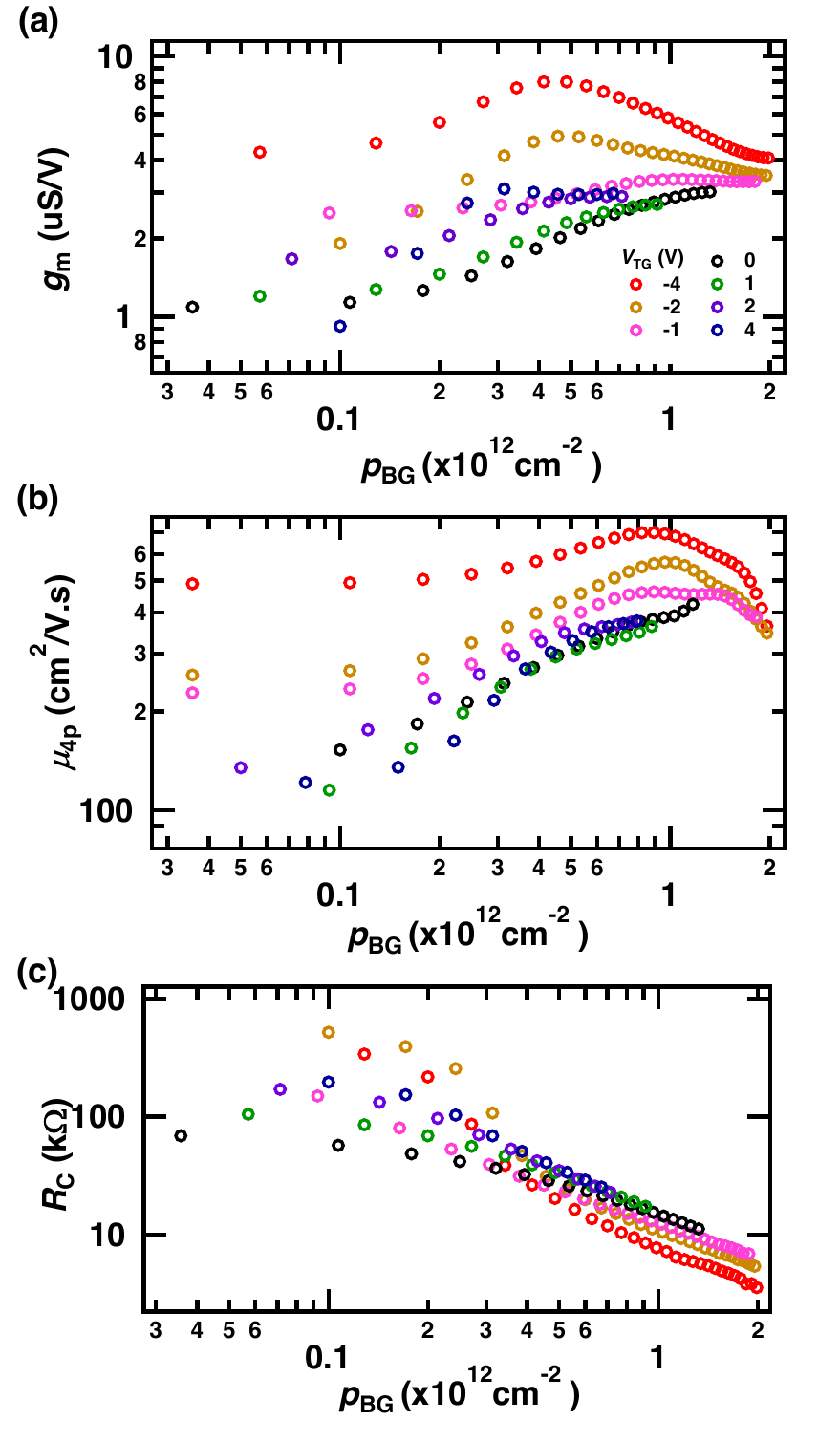}
       \caption{ a) The measured two-terminal back gate transconductance $g_{m}$ of the device shown in Figs.1 and 2 as a function of hole density $p_{BG}$ induced by the back gate, at fixed top gate voltages $V_{TG}$ at $T=$300 K. A four-fold enhancement of two-terminal $g_{m}$ is observed as $V_{TG}$ is tuned from 0~V to -4~V. b) The field effect mobility $\mu_{4p}$ of the same device on a log-log scale versus $p_{BG}$ at fixed top gate voltages at $T=$300~K. A two-fold enhancement in mobility is achieved by tuning top gate potential.  c) The contact resistance $R_{C}$ of the same device versus $p_{BG}$ at fixed top gate voltage. A two-fold modulation in contact resistance is observed as top gate voltage is tuned. }
       \label{Fig4}
\end{figure}

The transistor parameters of the bP FET were investigated in greater detail at $T$=300~K. Fig.~\ref{Fig4} (a) shows the two-terminal back gate transconductance $g_m = \partial I_{SD} /  \partial V_{BG} $ plotted versus the mobile hole density $p_{BG} = C_{BG} ( V_{BG} - V_{Th} ) / e$ induced by the back gate voltage, with a threshold voltage $V_{Th}$ that is dependent upon top gate voltage, and back gate capacitance $C_{BG} = 11.5\mathrm{nFcm}^{-2}$. The top-gate voltage strongly modulates the back-gate transconductance at fixed hole density. We measured the 4-point conductance $G_{xx}$ in our multi-terminal bP FET. The field effect mobility extracted from 4-point conductance $\mu_{4p} = \partial G_{xx} / \partial (C_{BG} V_{BG} )$ is plotted in Fig.~\ref{Fig4} (b) versus the induced hole density $p_{BG}$.

At low hole densities $p_{BG} < 4\times10^{11}\mathrm{cm}^{-2}$, the mobility increases as expected from the onset of percolation in the vicinity of the conduction threshold. At high hole densities, $p_{BG} > 8\times10^{11}\mathrm{cm}^{-2}$, the hole mobility falls with increasing carrier density, consistent with surface roughness scattering \cite{Kawaguchi, Ando}. The hole mobility is also modulated up to two-fold by the top gate voltage, with maximum mobility reached at $V_{TG}=-4V$, the most negative top gate voltage applied in our experiments. From our Schr\"{o}dinger-Poisson calculations at comparable hole density, we can infer that a negative top gate potential induces a hole accumulation layer at the top of the bP layer and that the volumetric hole density is spread through-out the bP layer. The hole accumulation layer induced at the top of the bP layer may contribute to the screening of trapped charge, reducing charged impurity scattering and enhancing mobility for holes within the bulk of the bP. The screening of trapped charge and concomitant increase of carrier mobility has been previously observed in bP by introduction of a graphene layer in close proximity to the bP layer, by which means significantly enhanced bP hole mobility has been observed \cite{Li_QHEbP}.

From the sample geometry and the combined measurement of two-point conductance $G_{2p}$ and four-point conductance $G_{xx}$, the contact resistance $R_C$ was determined and is plotted in Fig.~\ref{Fig4}(c) versus mobile hole density $p_{BG}$. As anticipated, the contact resistance to the hole gas within the bP layer decreases monotonically as the hole density within the bP layer increases. In addition to this expected trend, the top-gate potential is found to be effective at modulating the contact resistance at fixed hole density. The top gate electrode is ideally place for efficient electrostatic coupling to the region of carrier injection from contact electrode to the bP layer, as seen in Fig.~1. Our measurements of contact resistance is in accord with the electrostatically gated Schottky barrier model that has been successfully employed in the study of carbon nanotubes \cite{Heinze}, ultra-thin body silicon FETs and bP FETs \cite{Penumatcha} in two-point geometry. Unlike these previous studies, the effect of charge carrier distribution upon carrier injection into a low-dimensional FET channel is directly observed. 

The observation of mobility modulation effects in dual gate bP FETs demonstrates the capacity for bP to function as a room temperature VMT. Moreover, charge density distribution is seen to play an important role in the charge transport properties of 2D atomic crystals. The exposed surfaces of naked quantum well structures derived from 2D atomic crystals can lead to a strong spatial dependence of charge carrier scattering rates. In the specific case of bP, the bandgap range accessible by quantum confinement is ideal for applications in electronics, thermoelectrics and opto-electronics \cite{Gomez}. Band gap tuning of bP by a giant Stark effect \cite{kim} and by hydrostatic pressure \cite{Morita, xiang} have also been demonstrated, leading to a transition from direct gap semiconductor to Dirac semimetal in the extreme limit. The engineering of charge carrier distribution and confinement by externally applied potentials within thin bP layers adds a new means by which to tune and design bP quantum well device properties.

\section{Methods}
The bP crystals were 99.998\% purity from Smart Elements (Vienna, Austria). Nanometer-scale bP crystals were first exfoliated using a polydimethylsiloxane (PDMS) stamp technique \cite{Tayari_NC2015}. The exfoliation was performed inside a nitrogen glove box with O$_2$ and H$_2$O concentration below 5 ppm.  The thin bP crystals were transferred to degenerately doped Si/SiO$_{2}$ wafers that were previously dehydrated at $T=150^{\circ}$ C under vacuum and functionalized with a hexamethyldisilazane (HMDS) layer. Electrodes contacting the bP were defined using standard electron beam lithography (EBL) followed by 5~nm Ti/ 80~nm Au metal deposition. A top gate dielectric layer of 25~nm Al$_{2}$O$_{3}$ was deposited atop the bP by atomic layer deposition (ALD) at 150~$^\circ$C through an EBL defined window. A top gate metal layer was defined by a further EBL step followed by metal deposition (5nm Ti/ 80 nm Au). A final encapsulation by 300~nm of copolymer (methyl methacrylate) and 200~nm of polymer (polymethyl methacrylate) was performed at the end fabrication process.

Charge transport measurements were performed using quasi-dc excitation with a semiconductor parameter analyzer and vacuum probe station ($P\sim10^{-4}$ Torr) at room temperature.  A standard ac lock-in measurement technique was also used to measure FET conductance at a bias current $I_{SD}=1$~$\mu$A and a frequency $f=13.013$~Hz at $T=77$~K in a liquid nitrogen cryostat. After all electronic characterization, described further below, the encapsulating polymer layers were removed with warm acetone and atomic force microscopy (AFM) was performed within a glove box environment. 

\section{Acknowledgements}
The authors thank R. Martel, I. Fakih, D. Laroche for useful discussions, and Charlotte Allard for assistance with atomic force microscopy. Support was provided by the Natural Sciences and Engineering Research Council of Canada, the Fonds de Recherche du Qu\'ebec - Nature et Technologies, the Canada Research Chairs Program, and the Canadian Institute for Advanced Research. A portion of this work was performed at the National High Magnetic Field Laboratory which is supported by NSF Cooperative Agreement No. DMR-0084173, the State of Florida, and the DOE.

\end{document}